\documentclass[%
 reprint,
superscriptaddress,
 amsmath,amssymb,
 aps,
prb,
floatfix,
]{revtex4-1}

\usepackage{graphicx}% Include figure files
\usepackage{dcolumn}% Align table columns on decimal point
\usepackage{bm}% bold math
\usepackage{gensymb}
\usepackage{natbib}
\usepackage{url}
\usepackage{ctable}
\usepackage{multirow}
\usepackage{afterpage}
\usepackage{dcolumn}
\usepackage{amssymb}
\usepackage{textcomp}
\usepackage{xfrac}

\begin{document}

%\preprint{APS/123-QED}

\title[Machine learning potential]{Artificial neural network  interatomic potential for dislocation and fracture properties of Molybdenum}% Force line 

\author{Masud Alam}
 %\email{alam@mpie.de}
  \email{masud.alam@ruhr-uni-bochum.de}
 \author{Liverios Lymperakis}
  \affiliation{
Department of Computational Material Design, Max-Planck-Institut f{\"u}r Eisenforschung GmbH, Max-Planck Str. 1, D-40237 D{\"u}sseldorf, Germany}
% \email{Lymperakis@mpie.de}

\date{\today}% It is always \today, today,

\begin{abstract}

A high dimensional artificial neural network interatomic potential for Mo is developed. To train and validate the potential density functional theory calculations on structures and properties that correlate to fracture, such as elastic constants, surface energies, generalized stacking fault energies, and surface decohesion energies, have been employed. The potential provides total energies with a root mean square error less than 5\;meV per atom both in the training and validation data sets. The potential was applied to investigate screw dislocation core properties as well as to conduct large scale fracture simulations. These calculations revealed that the 1/2$\langle111\rangle$ screw dislocation core is non-degenerate and symmetric and mode I fracture is brittle. It is anticipated that the thus constructed potential is well suited to be applied in large scale atomistic calculations of plasticity and fracture.  

\end{abstract}

\maketitle

\section{\label{sec:level1}Introduction}

Plasticity and fracture of materials are governed by phenomena that span different length scales that range from a few $\mathrm{\AA}$, such as the dislocation core properties that govern intrinsic ductility and bond breaking at a crack tip, to the sub-$\mu$m regime such as long range strain interactions between pre-existing cracks and dislocations that control materials' failure  ~\cite{weygand2015multiscale, groh2009multiscale,horstemeyer2009multiscale,groh2009advances,de2007multiscale,curtin2003atomistic}. 
 Density functional theory (DFT) calculations can accurately describe the aforementioned phenomena in all relevant length scales. However, the computational cost still restricts first principles calculations to system sizes that do not exceed a few thousand atoms. As an alternative to DFT, semi-empirical potentials such as embedded atom method  (EAM)~\cite{daw1993embedded,daw1984embedded} and modified embedded atom method potentials (MEAM)~\cite{PhysRevB.68.144112,PhysRevB.46.2727,PhysRevB.64.184102,ahmad2018modified,groh2016modified}, bond order potentials~\cite{mrovec2004bond,pastewka2012bond,PhysRevB.69.094115}, and tight binding potentials~\cite{broughton1999concurrent,PhysRevLett.91.025501} are employed nowadays in atomistic calculations to investigate plasticity and fracture. These approaches allow to calculate systems consisting of a few million atoms and hence simultaneously describe phenomena at different length scales.
 
 A common pitfall associated with semi-empirical potentials is their transferability, i.e., their ability to accurately describe a wide range of material properties which were not employed in the potential parametrization. For example, to model fracture, a potential should be able to accurately describe the structural and elastic properties of the material as well as the highly strained region of dislocation cores, surface and unstable stacking faults energies~\cite{groh2015fracture,rice1992dislocation,griffith1921vi}, and surface decohesion~\cite{ko2014origin,wu2015magnesium}.

 Machine learning (ML) potentials have attracted considerable interest in computational materials science. The idea underlying ML potentials is to bypass the physical insights into the potential functional form. Instead, machine learning methods are applied to fit the potential energy surface and construct a functional form that provides the energy of each atomic configuration. An advantage of this approach is that the number of potential parameters is not limited and thus rich training sets that include a wide range of material properties can be fitted~\cite{behler2021machine,behler2016perspective,zuo2020performance,PhysRevMaterials.1.043603}.

 ML potentials have been developed and successfully applied to investigate plasticity and fracture properties. For example, a gaussian approximation potential (GAP) has been developed and applied to study the structure and mobility of screw dislocations in bcc Fe~\cite{maresca2018screw}. It was revealed that the GAP was able to address artifacts encountered by traditional semi-empirical approaches such as the split/degenerate dislocation and calculate a large variety of properties in agreement with DFT calculations. The properties of screw dislocations in bcc Fe have been also investigated by an artificial neural network (ANN) atomic potential~\cite{mori2020neural}. It was shown that the ANN potential was able to calculate dislocation properties and dynamics in excellent agreement with the DFT calculations. ANN potentials have also been developed and applied to study dislocations and fracture in Al and Mg~\cite{stricker2020machine,PhysRevMaterials.4.103601}.  
These potentials revealed an excellent agreement to DFT calculated dislocations core properties for many slip systems. 
Moreover, the brittle-ductile transition was also well captured for different crack orientations. 
 
The aim of the present study is to train and evaluate a ML potential for plastic and fracture atomistic simulations. We, therefore, focus on the dislocation core and fracture properties of the bcc transition metal Mo. Mo is primarily used as an alloying element with the steel to increase the strength and hardness of the alloy as well as to increase the resistance to corrosion~\cite{liu2013nanostructured, tabernig2010joining,braithwaite2013molybdenum}. 
Until now many ML potentials have been developed for Mo. These include four major types of interatomic potentials: The spectral neighbor analysis potential, the gaussian approximation potential, the moment tensor
potentials, and neural network potential ~\cite{PhysRevMaterials.1.043603,doi:10.1021/acs.jpca.9b08723}. The performance of these potentials has been tested on a wide range of technologically important properties such as elastic constants, surface, and grain boundary energies, generalized stacking fault energies, diffusion barriers, and melting points. These reports clearly demonstrate that ML potentials are able to achieve near DFT accuracy in predicting material properties and outperform existing "traditional" EAM and MEAM type potentials.

In the present study, we trained a Behler-Parinello type ANN potential~\cite{PhysRevLett.98.146401}.
As the low-temperature plasticity of bcc materials is controlled by the properties of the 1/2$\langle111\rangle$\ screw dislocations, the core properties of this defect have been addressed by DFT and interatomic potential calculations~\cite{PhysRevB.85.214121,PhysRevB.89.024104,ismail2000ab}.
Dezerald~\textit{et al.} employed DFT calculations and reported that the core structure of the 1/2$\langle111\rangle$\ screw dislocation is non-degenerate and the easy core structure is the lowest energy configuration~\cite{dezerald2014ab}. Non-degenerate core structures have been also reported by MEAM~\cite{PhysRevB.85.214121} and bond order potential calculations~\cite{mrovec2004bond}. These calculations revealed a symmetric core with the displacement field spread in the \{110\} planes. However, other Mo interatomic potentials of Mo yield degenerate cores: For example, Finnis-Sinclair potential predicts a degenerate screw dislocation core~\cite{mendis2006use, li2004core}.

Although atomistic investigations of fracture in Mo are scarce, a comparative study of semi-empirical potentials of bcc Fe indicated that potentials sharing the same functional form and/or yielding similar surface energies and unstable stacking fault energies predict different crack tip phenomena such as structural transformation, crack kinking, presence of planar faults~\cite{moller2014comparative} etc. Moreover, an often encountered artifact of interatomic potentials in fracture simulations is the crack tip blunting without dislocation emission and the artificial oscillations in surface decohesion energy profiles at large separation distances~\cite{groh2015fracture}. 

In order to train and evaluate the potential, a material's database has been constructed which consists of material properties that are related to dislocation and fracture such as the generalized stacking fault energies, surface energies, surface decohesion energy curves, elastic constants, etc. These have been calculated by employing DFT calculations. The atomic structures and the total energies of the above DFT calculations were then used to train the neural network. Employing the newly derived ANN potential, the properties of the $1/2\langle 111 \rangle$ screw dislocation core have been investigated. The fracture properties have been evaluated in terms of linear elastic fracture mechanics~\cite{rice1992dislocation,griffith1921vi} as well as by employing large scale atomistic fracture simulations.   

The article is organized as follows: In section~\ref{sec:method} the details of the NN architecture and as well as of the \textit{ab-initio} calculations are given. 
In Section~\ref{sec:result}, we present the application of the ANN potential to calculate the dislocation core properties. These have been evaluated in terms of the Nye tensor~\cite{hartley2005representation} and the differentials displacement maps~\cite{vitek1968intrinsic}.
In Section~\ref{sec:frac} the atomistic calculations of fracture at a crack tip in Mo are presented. These are compared to predictions by linear elastic fracture mechanics.
In section~\ref{conclu}, we summarize.
	
\section{Methodology}\label{sec:method}

\subsection{\label{sec:dft}\textit{Ab-initio} calculations}

In order to build the materials' database, we conducted first principle calculations within the spin-polarized density functional theory (DFT). The calculations have been performed employing the Vienna \textit{ab-initio} simulation package, VASP~\cite{kresse1996software,PhysRevB.50.17953}. A kinetic energy cutoff of 400\;eV was used for the expansion of the plane-wave basis set. An equivalent $6\times6\times6$ Monkhorst-Pack $k$-point mesh  for the bulk unit cell was used to sample the Brillouin zone and the generalized gradient approximation (GGA)~\cite{perdew1996generalized} was employed for the exchange and correlation.
The positions of the atoms have been relaxed until the forces on each atom are less than 0.01\,$\mathrm{eV}/\mathrm{\AA}$. The total energy and force calculations with the ANN, the EAM, and MEAM potentials have been performed using the Large-scale Atomic/Molecular Massively Parallel Simulator (LAMMPS)~\cite{plimpton2007lammps}. Unless otherwise stated, the atomic relaxations have been stopped until either the change in the total energy was lower than $10^{-6}$\;eV or the forces on each atom were less than $10^{-6}$\;$\mathrm{eV}/\mathrm{\AA}$.

\subsection{\label{sec:interpot}Machine Learning potentials}

\begin{figure}[t]
	\includegraphics[width=\columnwidth]{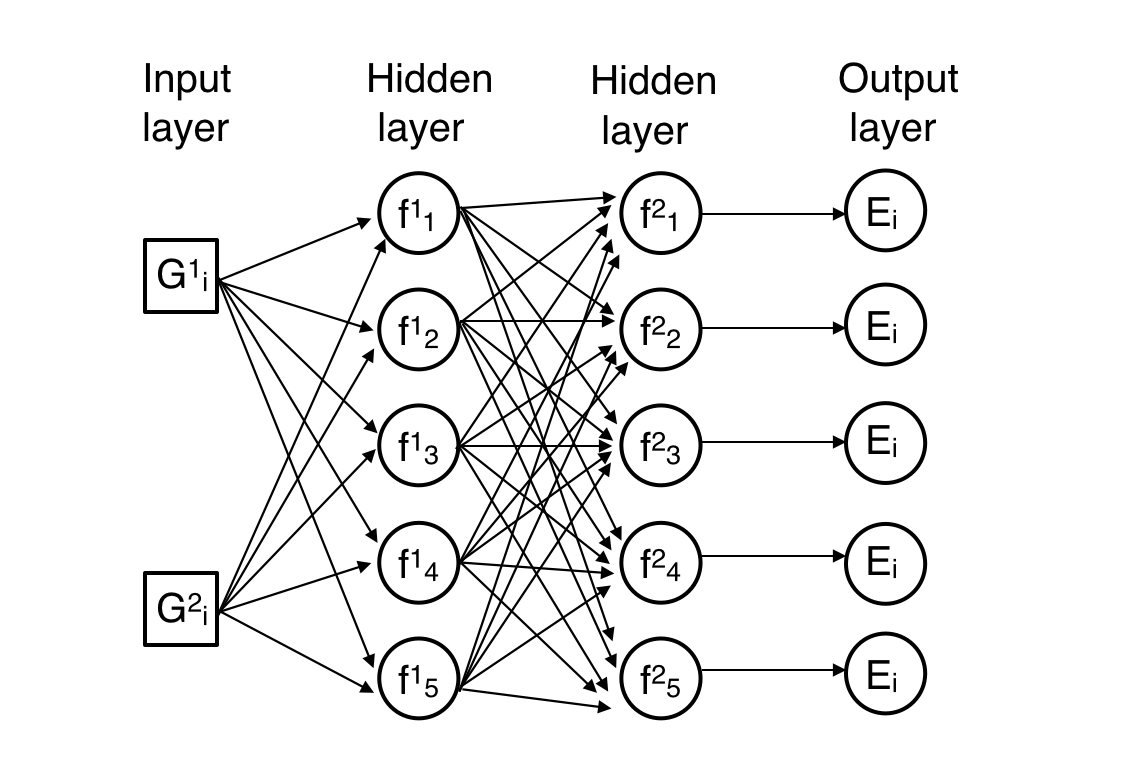}
	\caption{Schematic representation of an ANN architecture consisting of two hidden layers and one output layer. The input layer consists of radial and angular symmetry functions. }
	\label{fig:archi}
\end{figure}

\begin{table}
\caption{Parameters of the 15 radial  $G_i^1$ and 18 angular $G_i^2$ symmetry functions. The cutoff distance, $r_c$ is set to 6.5~$\mathrm{\AA}$ and the centre of the Gaussians, $r_s$ is set to 0 in all  functions. The units of $\eta_a$ and $\eta_s$ are $\mathrm{\AA}^{-2}$}
\begin{ruledtabular}
\begin{tabular}{c | c | ccc}
No. & Radial & \multicolumn{3}{c}{Angular} \\
\hline
 &$\eta_s$ &$\eta_a$ &$\lambda$ &$\zeta$ \\
\hline
 1 & 0.005025 &  0.005025 & -1 & 1 \\
 2 &0.105136 & 0.005025 & -1 & 2\\
 3 & 0.205147 & 0.005025 & -1 & 4\\
 4 &0.305059  &0.005025 &1  &1 \\
 5 &0.404871  &0.005025 &1  & 2\\
 6 &0.504583  &0.005025 &1  &4 \\
 7 &0.604197  &0.237571 &-1  &1 \\
 8 &0.703711  &0.237571 &-1  & 2\\
 9 &0.803126  &0.237571 &-1  & 4\\
 10 &0.902441  &0.237571 &1  &1 \\
 11 &1.001658  &0.237571 &1  &2 \\
 12 &1.100776  &0.237571 &1  &4 \\
 13 &1.199796  &0.469578 &-1  &1 \\
 14 &1.298717  &0.469578 &-1  &2 \\
 15 &1.397539  &0.469578 &-1  & 4\\
 16 &  &0.469578 &1  & 1\\
 17 &  &0.469578 &1  &2 \\
 18 &  &0.469578 &1  &4 \\
		\end{tabular}
	\end{ruledtabular}
\end{table}

\begin{figure}[t]
	\includegraphics[width=\columnwidth]{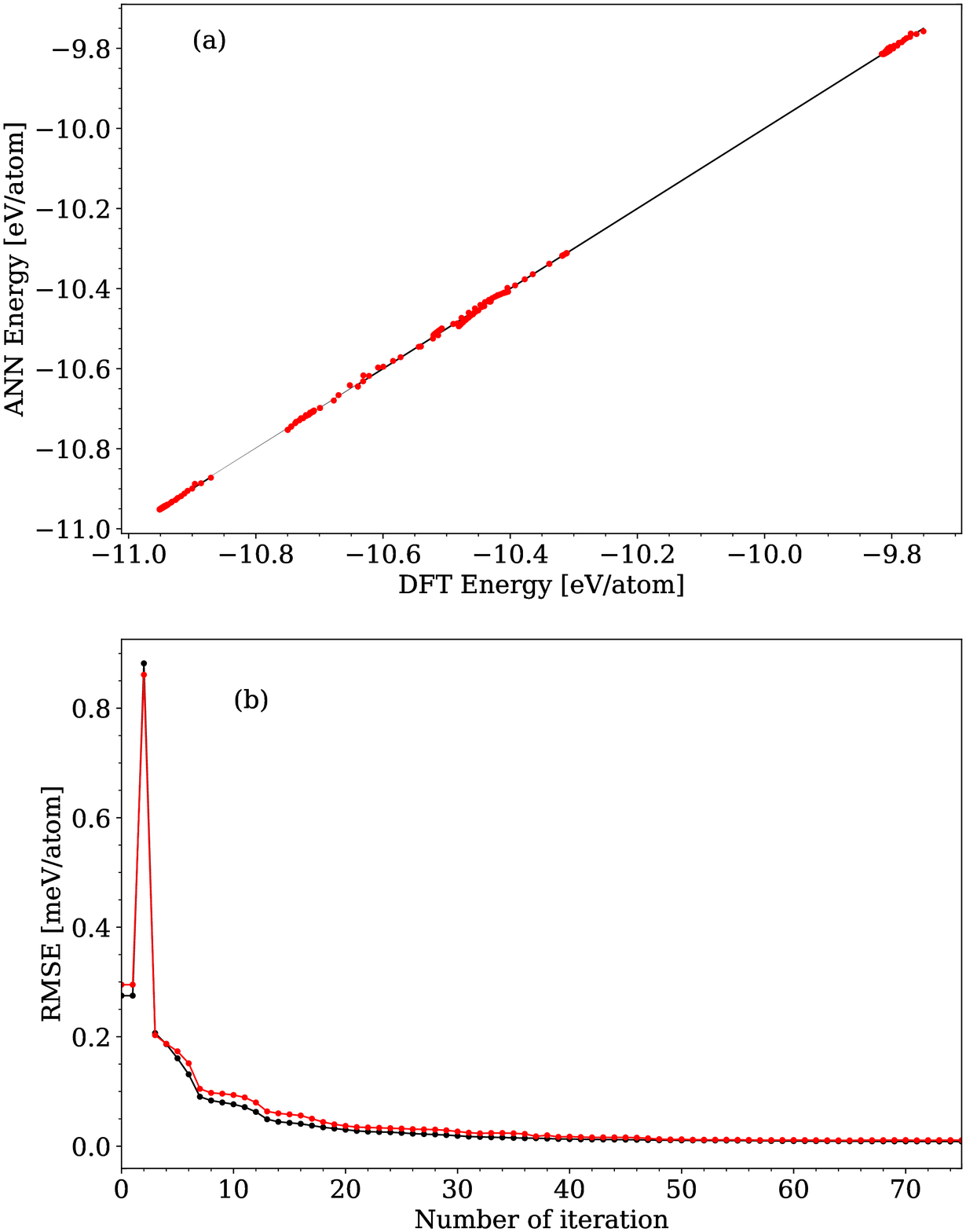}
	\caption{(a) ANN vs DFT calculated energies per atom of all structures in the material's database. The solid line represents the perfect match between ANN and DFT calculations. (b) Evolution of root mean square error as a function of the number of iterations. Black dots represent the error in the training set, and red dots in the testing set. For the sake of clarity only the first 75 out of 200 iterations are shown.}
	\label{mll}
\end{figure}
For the ANN ML potential, we have adopted the formalism by Behler and Parrinello~\cite{PhysRevLett.98.146401} and the atomic energy network (aenet) software package was used for train and evaluation~\cite{artrith2016implementation,PhysRevMaterials.4.040601}. This formalism implements a feed forward NN and the energy of the system is the sum of all the atomic energies $E_i$, i.e., $E=\sum_i^{N} E_i$, where $E$ is the total energy and $N$ is the number of atoms. Each atomic energy is a function of the local structural environment and it is the output of a node at the outer layer of the NN (see Fig.~\ref{fig:archi} for a schematic representation). For the case of the ANN shown in Fig.~\ref{fig:archi}, $E_i$ is written as
\begin{equation}
E_i = f_a^2\left[w_{01}^2+\sum_{j=1}^3w_{j1}^2f_a^1\left(w_{0j}^1+\sum_{\mu=1}^2w_{\mu j}^1G_i^{\mu}\right)\right],
\label{mleqn}
\end{equation}
where, $w_{ij}^k$ is the weight parameter connecting node $j$ in layer $k$ with node $i$ in $k-1$ layer. Furthermore, all but the nodes in the input layer, are connected to a bias node with  $w_{0j}^{k}$ weight parameter. $f_a^k$ is the activation function of layer $k$ and node $a$. The hyperbolic tangent function has been chosen for the activation function of the hidden layers and a linear function for the activation function of the output layer.

$G_i^{\mu}$ is the symmetry function of atom $i$ and of type $\mu$. These functions depend on the position of all atoms within a cutoff radius $r_c$ and are invariant to translations, rotations as well as to exchange of any two atoms of the same chemical species.
The right choice of the symmetry function is the most important step in generating ANN potentials. In the present study, we have adopted atom centered basis set for the generating symmetry function~\cite{artrith2016implementation,behler2011atomnn}. Similar method has been applied to construct the symmetric function of Al-Cu, Mg, Al-Mg-Si ANN ML potentials~\cite{PhysRevMaterials.4.103601,PhysRevMaterials.1.053604,artrith2016implementation,stricker2020machine}. 

In the present work we used radial, denoted as $G_i^1$, and angular, denoted as $G_i^2$, symmetry functions. The radial functions contain the sum of two body terms while in the angular functions three body terms are also included. 
For the radial symmetry function, we consider the sum of Gaussians which are smoothly truncated at a distance $r_c$ by the application of a cutofff function $f_c$. Hence, $G_i^{1}$ is written as
\begin{equation}\label{g2}
G_i^{1}=\sum_{i\ne j}^{N} e^{-\eta_s\left(r_{ij}-r_s\right)^2}f_c\left(r_{ij}\right),
\end{equation}
where $r_{ij}$ is the interatomic distance between atoms $i$ and $j$, and $\eta_s$ and $r_s$ are the widths and the centre of the Gaussian respectively. 

The cutoff function $f_c(r_{ij})$ smoothly truncates the symmetry function at the cutoff distance $r_c$ . The functional form of $f_c(r_{ij})$ is chosen such that its second derivative with respect to position is everywhere continuous~\cite{artrith2016implementation}:
\begin{equation}\label{cuteq}
f_c(r_{ij})=
\begin{cases}
0.5\left[cos\left(\frac{\pi\;r_{ij}}{r_c}\right)+1\right] & \text{ if }   r_{ij} \leq r_c \\
0  &  \text{ if } r_{ij} > r_c
\end{cases}
\end{equation}
The cutoff distance, $r_c$ was set to 6.5~$\mathrm{\AA}$ in order to include second nearest neighbor interactions in the potential.

For the angular symmetry functions we used the following formalism:
\begin{align}\label{g4}
G_i^{2}=2^{1-\zeta}\sum_{j,k \ne i}^N \left(1+\lambda \;cos\;\theta_{ijk}\right)^{\zeta}\;e^{-\eta_a(r_{ij}^2+r_{ik}^2+r_{jk}^2)}\;\\
 \nonumber f_c(r_{ij})f_c(r_{ik})f_c(r_{jk}).
\end{align}
Here, $\theta_{ijk}$ is the angle between $\boldsymbol{r_{ij}}$ and $\boldsymbol{r_{ik}}$. The parameter $\lambda$ can attain two values, +1 and -1, which shift the maxima of the cosine to $0^{\circ}$ and to $180^{\circ}$, respectively. 
The parameter, $\zeta$ determines the range of angles of finite function values. As in Eq.~(\ref{g2}), $\eta_{a}$ is the width of the Gaussian.

The present NN architecture consists of 4 layers. In the first layer, there are 33 nodes that correspond to the 15 and 18 radial and angular symmetry functions, respectively.
We have constructed the symmetry function in such a way that atoms having different symmetry yield different energies, i.e, a pair of symmetry functions for which the difference in their values $|\Delta G| \rightarrow 0$ but the difference in the output atomic energies $|\Delta E| >> 0$ are excluded~\cite{behler2011atomnn}.
This allows for faster convergence of the force and energy components.
The complete network contains in total 1121 weights ($w_{ij}^{k}$) including bias.

To build the ANN potential, we have considered 326 atomic structures. 32 structures were chosen randomly for testing and 294 structures for training. These include Mo in bcc, fcc, hcp, and sc phases at various hydrostatic strains and Mo in bcc and sc at various hydrostatic, uniaxial and biaxial strain states. Furthermore, the $\{100\}$, $\{110\}$ surface energies of bcc Mo, the  generalized stacking fault energies of the $\{112\}$  and $\{100\}$ planes for displacement along $\langle111\rangle$, and the traction separation energies of the $\{110\}$ plane have been included in the structure-energy dataset. The optimization of the weights ($w_{ij}^{k}$) was performed using the Broyden-Fletcher-Goldfarb-Shanno algorithm (BFGS) with 200 iterations. The above NN architecture yielded a root mean square error (RMSE) smaller than 5\;meV/atom both for training and testing sets, as can be seen in Fig.~\ref{mll}.

\begin{table*}[t]
	\caption{Calculated properties of Mo using the present ANN potential as well as MEAM~\cite{PhysRevB.85.214121} and EAM~\cite{PhysRevB.69.144113} potentials and DFT. All the properties are for bcc Mo unless it is otherwise denoted. $\Delta E$ denotes the difference in the cohesive energy in meV/atom of other Mo crystalline phases with respect to the bcc structure. $a$ is the lattice constant in $\mathrm{\AA}$ and $c/a$ the ratio of lattice constants of the hcp structure. The unit of the bulk modulus is GPa. The units of the other properties are listed in the Table. The superscript p ($^\mathrm{p}$) indicates present DFT calculations, and $^\mathrm{(*)}$ indicates properties that were included into the training dataset.
	}\label{Tabni}
	\begin{ruledtabular}
		\begin{tabular}{l*{5}{c}r}
			&&ANN &MEAM~\cite{PhysRevB.85.214121} &EAM~\cite{PhysRevB.69.144113}   & DFT \\
			\hline 
			\multirow{3}*{bcc Mo} &$^\mathrm{(*)}a$&3.15& 3.167&3.15 &3.15$^\mathrm{p}$, 3.169~\cite{PhysRevB.85.214121}       \\  
			&$^\mathrm{(*)}$B &270&257    &257    &269$^\mathrm{p}$, 263~\cite{PhysRevB.85.214121}     \\
			&$^\mathrm{(*)}$B$^{'}$           &4.67&5.06  &4.69  &4.51$^\mathrm{p}$,       \\
			\hline 
			\multirow{2}*{fcc}  &$^\mathrm{(*)}$$\Delta E$&435&392 &154    &434$^\mathrm{p}$, 418~\cite{PhysRevB.85.214121}     \\   
			&$^\mathrm{(*)}$$a$&3.99&3.93&4.12  & 3.99$^\mathrm{p}$,  4.013~\cite{PhysRevB.85.214121}    \\ 
						\hline 
			\multirow{2}*{hcp} &$^\mathrm{(*)}$$\Delta E$&455&415&154&470$^\mathrm{p}$, 433~\cite{PhysRevB.85.214121}   \\   
			&$^\mathrm{(*)}$$a$&2.82&2.79 &2.91    &2.80$^\mathrm{p}$, 2.765~\cite{PhysRevB.85.214121}   \\  
			&$^\mathrm{(*)}$$c/a$&1.63 &1.633 &1.633    &1.67$^\mathrm{p}$, 1.77~\cite{PhysRevB.85.214121}   \\ 
			\hline 
			\multirow{2}*{sc } &$^\mathrm{(*)}$$\Delta E$&1139&3429 &1756    &1139$^\mathrm{p}$   \\   
			&$^\mathrm{(*)}$$a$&2.59&3.16&2.44&2.59$^\mathrm{p}$       \\  
			\hline 
			\multirow{2}*{dia } &$\Delta E$&1497&2893 &2885    &2198$^\mathrm{p}$    \\   
			&$a$&5.68&5.98&5.98&5.64$^\mathrm{p}$       \\  
			\hline 
			\multirow{3}*{Elastic constant (GPa) }
			&$^\mathrm{(*)}$C$_{11}$  &473  &441 &456 &492$^\mathrm{p}$, 462 \cite{PhysRevB.85.214121}   \\   
			&$^\mathrm{(*)}$C$_{12} $  &178&158&167  &155$^\mathrm{p}$, 163 \cite{PhysRevB.85.214121}       \\   
			&$^\mathrm{(*)}$C$_{44}$   &115&96&113  &104$^\mathrm{p}$, 102 \cite{PhysRevB.85.214121}      \\   
			\hline
			\multirow{1}*{Vacancy (eV)}
			&$E_{v}$ &2.61&2.96&2.95& 2.79~\cite{PhysRevB.85.214121}     \\              
			\hline
			\multirow{3}*{Surface (meV/\AA{}$^2$)}
			&$^\mathrm{(*)}$$\gamma_{(100)}$ &194 &180&154      &203$^\mathrm{p}$, 200~\cite{PhysRevB.85.214121}     \\  
			&$^\mathrm{(*)}$$\gamma_{(110)}$ &168&164&134       &175$^\mathrm{p}$, 174~\cite{PhysRevB.85.214121}  \\
			&$\gamma_{(111)}$ &183&201&172       &183$^\mathrm{p}$, 186~\cite{PhysRevB.85.214121} \\
			\hline
			\multirow{2}*{Unstable stacking fault energy (meV/$\mathrm{\AA}^2$)}
			&$U_{(110)}$ &70.52 &91.63&89.80& $78.33^\mathrm{p}$    \\  
			&$^\mathrm{(*)}$$U_{(112)}$  &90.43  &98.36&103.94& $85.22^\mathrm{p}$ \\
		\end{tabular}
	\end{ruledtabular}
\end{table*}

\subsection{\label{sec:otherprop} Structural and Elastic Properties}

In the materials' database to train the ANN potential, we have included the energetics of bcc, fcc, hcp, and sc Mo at various strain states as well as the equilibrium lattice and elastic constants. The results are shown in Table~\ref{Tabni}.
 The cohesive energy difference of the low energy polymorphs are in excellent agreement with the DFT calculations. The ANN potential calculated cohesive energy differences of fcc, hcp, and sc with respect to the bcc phase are 435, 455, and 1139 meV/atom, respectively. The DFT calculated values are 434, 470, and 1139 meV, respectively. 
Nevertheless, the largest deviation is for the diamond structure. The DFT predicted energy difference of diamond Mo to the ground state bcc is 2198 meV/atom, whereas the ANN potential yields an energy difference of 1497 meV/atom.

The ANN potential calculated lattice constants are in excellent agreement with the DFT calculated values (less than 2\% difference). The bulk modulus of bcc Mo is obtained by fitting the energy vs volume data to the Murnaghan equation of state~\cite{murnaghan1944compressibility} and it is in excellent agreement with the DFT calculated value (270 vs 269 GPa, respectively). 
We have also calculated the elastic constants of bcc Mo ($C_{11}$, $C_{12}$, and $C_{44}$).  
As can be seen in Table~\ref{Tabni}, the ANN potential calculated elastic constants are within less than 10\% in agreement with the DFT calculated values. 

We have further included in the database the vacancy formation energy in bcc Mo by employing a $4\times4\times4$ supercell. The vacancy formation energy, $E_v$ is calculated as
\begin{equation}
E_v=E_\mathrm{tot}^v\left(n\right)-n \mu_\mathrm{bulk}\;,
\end{equation}
where $E_\mathrm{tot}^v\left(n\right)$ is the total energy of the system consisting of $n$ atoms and a vacancy. $\mu_\mathrm{bulk}$ is the chemical potential of bulk bcc Mo.
The ANN potential calculated vacancy formation energy is 2.61\;eV which is in excellent agreement with the DFT value of 2.79\;eV.
%%%%%%%%%%%%%%%%%%%%%%%%
%%%%%%%%%%%%%%%%%%%%%%%%
%%%%%%%%%%%%%%%%%%%%%%%%
\subsection{\label{sec:SurfEn} Surface Energies}

The brittle fracture of materials is associated with plane cleavage. Hence, a necessary prerequisite for an interatomic potential to describe fracture is to accurately calculate surface energies~\cite{griffith1921vi}.
As can be seen in Table~\ref{Tabni}, both EAM and MEAM potentials fail to qualitatively describe the energetics of low index surfaces,
i.e., the predicted ordering of the energetically most favorable surfaces is not in agreement with the DFT calculations. 
Therefore, these potentials are prone to calculate inconsistent critical stress intensity factors with respect to DFT. 

We have included the energies of the $\{100\}$,  $\{110\}$, and $\{111\}$ surfaces in the materials' database. The surface energies have been calculated with the ANN potential using a slab geometry consisting of 20 unit cells along the normal direction to the surface, a vacuum region of 10\;$\mathrm{\AA}$, and $1\times1$ surface unit cell. 
The surface energy, $\gamma$, is defined as
\begin{equation}
\gamma=\frac{1}{2 A}\left(E_\mathrm{slab} - n E_\mathrm{bulk}\right)\qquad,
\label{eq:surfNRG}
\end{equation} 
where $E_\mathrm{slab}$ is the total energy of the slab consisting of $n$ atoms, and $E_\mathrm{bulk}$ is the total energy per atom of bulk. $A$ is the surface area and the factor 2 in the denominator accounts for the two symmetry equivalent surfaces in the slab geometry. 

The ANN potential calculated  energies of the $\{100\}$, $\{110\}$, and $\{111\}$ surfaces are 194, 168, and 183\;meV/$\mathrm{\AA}^2$, respectively whereas the DFT calculated energies are 203, 175, 183 meV/$\mathrm{\AA}^2$, respectively. Hence, the ANN potential provides an excellent qualitative agreement with DFT. Furthermore, the quantitative agreement to DFT is better than 9\;meV/$\mathrm{\AA}^2$.
%%%%%%%%%%%%%%%%%%%%%%%%
\subsection{\label{sec:deco} Surface decohesion}
\begin{figure}[t]
	\includegraphics[width=\columnwidth]{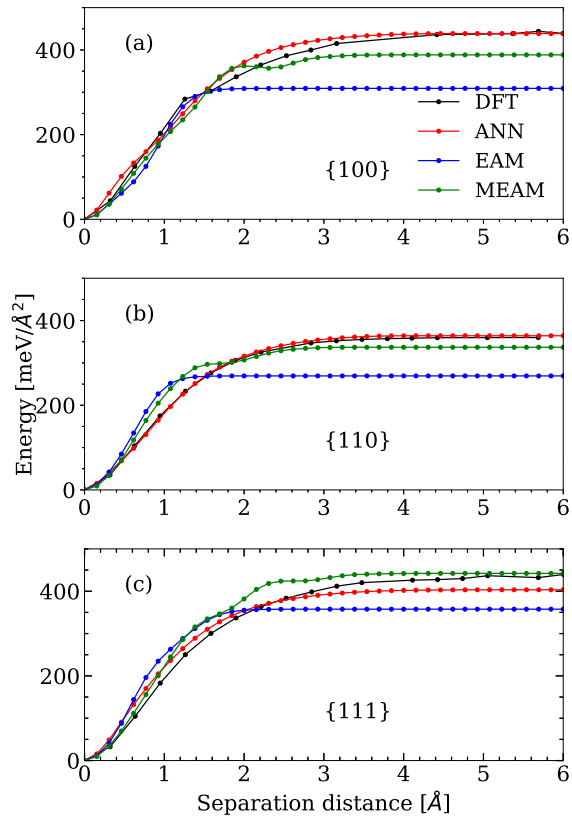}
	\caption{Separation energy of (a) $\{100\}$ (b) $\{110\}$ (c) $\{111\}$ surfaces of Mo as a function of separation distance as obtained with DFT, ANN, MEAM, and EAM potentials. 
	%$\{100\}$ and $\{110\}$ plane energy vs displacement was included in the fitting database whereas $\{111\}$ was used to validate the prediction of ANN potential. 
	}
	\label{trabre}
\end{figure}

A common problem with interatomic potential calculations of surface decohesion, i.e., the separation of the material across a surface, is the spurious oscillatory behavior of the energy profiles at large separation distances~\cite{alam2021meam,groh2015fracture,wu2015magnesium}. This oscillatory behavior is an artifact of the interatomic potential and stems from the abrupt truncation of the pair interactions at the cut-off distance of EAM and MEAM potentials. Moreover, it introduces a potential barrier for brittle-ductile transition and artificial crack blunting~\cite{ko2014origin}. Furthermore, it affects the crack tip opening displacement, the cohesive strength, and the work of separation~\cite{needleman1987continuum}. Therefore, surface decohesion represents a challenging testbed for interatomic potentials to model fracture.

In order to calculate the surface decohesion energy profiles, a supercell was created consisting of 20 units along the normal to the surface direction and  $1\times1$ surface cell. The cell was then split into two halves, and the upper half was incrementally and rigidly shifted upwards with a step of 0.15\;$\mathrm{\AA}$. The decohesion energy was calculated as 
\begin{equation}
E_d(\Delta z)=\frac{1}{A} (E_{\Delta z}-E_0)
\label{eq:dec}
\end{equation}
where $E_{d}( \Delta z )$ is the energy of the system at separation distance $\Delta z$, $E_0$ is the energy of the bulk, and $A$ is the area of the surface.

Fig.~\ref{trabre} shows the energy vs separation distance for three low index surfaces, i.e., the $\{100\}$, $\{110\}$, and $\{111\}$, calculated by DFT, ANN, MEAM, and EAM potentials. The decohesion energies are underestimated by the EAM potential while the MEAM potential shows oscillatory profiles. This introduces artificial repulsive forces between the two surfaces at regions of large separation distances. For large separation distances, the decohesion energy equals twice the unrelaxed surface energy [compare Eqs.~(\ref{eq:surfNRG}) and (\ref{eq:dec})]. Therefore, at large separation distances the deviations from DFT follow the same trends as the energies of the relaxed surfaces (see Table~\ref{Tabni}). The tensile stress and the cohesive strength are proportional to the slope and the maximum gradient of the decohesion energy curve, respectively. The EAM and MEAM potentials overestimate the slope for the decohesion of the \{110\} surface. On the other hand, the present ANN shows an excellent agreement with the DFT calculated profiles.  

%%%%%%%%%%%%%%%%%%%%%%%%
%%%%%%%%%%%%%%%%%%%%%%%%

\subsection{\label{sec:gsfe} Generalized Stacking Fault Energies}
\begin{figure}[t]
	\includegraphics[width=\columnwidth]{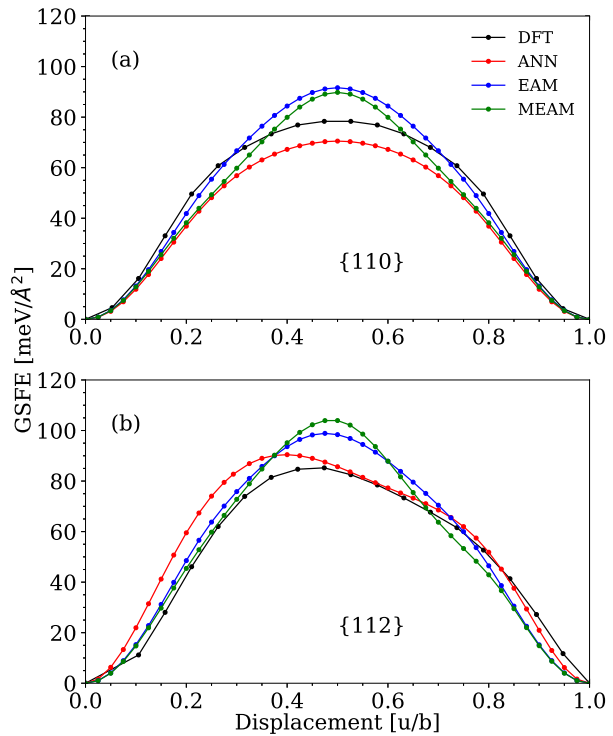}
	\caption{GSFE profiles for Mo as function of displacement along $\langle\bar{1}11\rangle$ direction for (a) $\{\bar{1}10\}$ and (b) $\{\bar{1}\bar{1}2\}$  plane obtained from DFT, ANN potentials, as well as from MEAM~\cite{PhysRevB.85.214121} and EAM~\cite{PhysRevB.69.144113} potentials. 
	}
	\label{gsf}
\end{figure}
The emission of a dislocation from a crack tip depends on the unstable stacking fault energy~\cite{rice1992dislocation}. 
For an edge dislocation, the dislocation core width is inversely proportional to unstable stacking fault energy and the Peierls stress decays exponentially with the dislocation core width~\cite{PhysRevB.50.5890,PhysRevLett.78.266}. The GSFE has been included in the materials database to fit and validate the ANN potential: The GSFE for sliding along $\langle111\rangle$\ in the $\{112\}$ plane was used to train the potential whereas that for sliding along $\langle111\rangle$\ in the $\{110\}$ plane was used for validation. 

In order to calculate the aforementioned GSFEs, slab geometries consisting of 20 unit cells along the normal to the surface direction and a vacuum region of 10\;$\mathrm{\AA}$ have been employed. The bottom half of the slab was kept fixed while the upper half was rigidly shifted along the sliding direction. 
The GSFE profile is calculated as
\begin{equation}
U_{x_i}=\frac{1}{A}\left(E_{x_i} - E_\mathrm{0}\right)\qquad,
\label{eq:gsfe}
\end{equation} 
where $U_{x_i}$ and $E_{x_i}$ are the GSFE and the total energy at displacement $x_i$, respectively. $E_0$ and $A$ are the energy of the system without applied displacement and the area of the interface, respectively.

Figs.~\ref{gsf}(a) and (b) show the GSFE against the normalized displacement for slip in the $\{110\}$ and $\{112\}$ planes, respectively. The DFT calculated asymmetric profile of the GSFE for slip in the $\{112\}$ plane is well described by the present ANN potential. The EAM and MEAM potentials on the other hand yield a more symmetric profile. Nevertheless, all potentials correctly reproduce the shape of the GSFE for slip in the $\{110\}$ plane. The MEAM and EAM potentials overestimate the unstable stacking fault energy by $\approx$\;15\% with respect to DFT, whereas the present ANN potential underestimates the energy by $\approx$\;9\% for the slip in the $\{110\}$ plane. As the unstable stacking fault energy of the $\{110\}$ plane is lower than that of the $\{112\}$ plane, the $\langle111\rangle$ $\{110\}$ slip system is foreseen to control slip over the $\langle111\rangle$ $\{112\}$.
%%%%%%%%%%%%%%%%%%%%%%%%
%%%%%%%%%%%%%%%%%%%%%%%%
%%%%%%%%%%%%%%%%%%%%%%%%
%%%%%%%%%%%%%%%%%%%%%%%%
%%%%%%%%%%%%%%%%%%%%%%%%
%%%%%%%%%%%%%%%%%%%%%%%%
\section{Dislocation Core }\label{sec:result}

Plastic deformation depends on the dislocation core properties such as the symmetry, the degeneracy, and the core width~\cite{heggen2010plastic}. 
Due to the high intrinsic lattice resistance, the low temperature plastic flow of bcc materials is controlled by the mobility of screw dislocations~\cite{hull2001introduction}. These dislocations are less mobile than their edge counterparts and move through double kink nucleation~\cite{li2004core,PhysRevB.85.214121,gordon2010screw}.
Nevertheless, often different interatomic potentials and/or different potential parametrizations for the same material predict different dislocation core properties. For example, it has been shown that EAM, bond order, and MEAM potentials predict degenerate and non-degenerate $\frac{1}{2}\langle111\rangle$ screw dislocation core in bcc W~\cite{cereceda2013assessment}. 
Furthermore, interatomic potentials commonly produce split screw dislocation core structures which influence the slip behavior of kinked dislocations~\cite{hale2014simulations}. Therefore, in the present study, we choose to validate our potentials by calculating the dislocation core properties.

\begin{figure}[t]
	\includegraphics[width=\columnwidth]{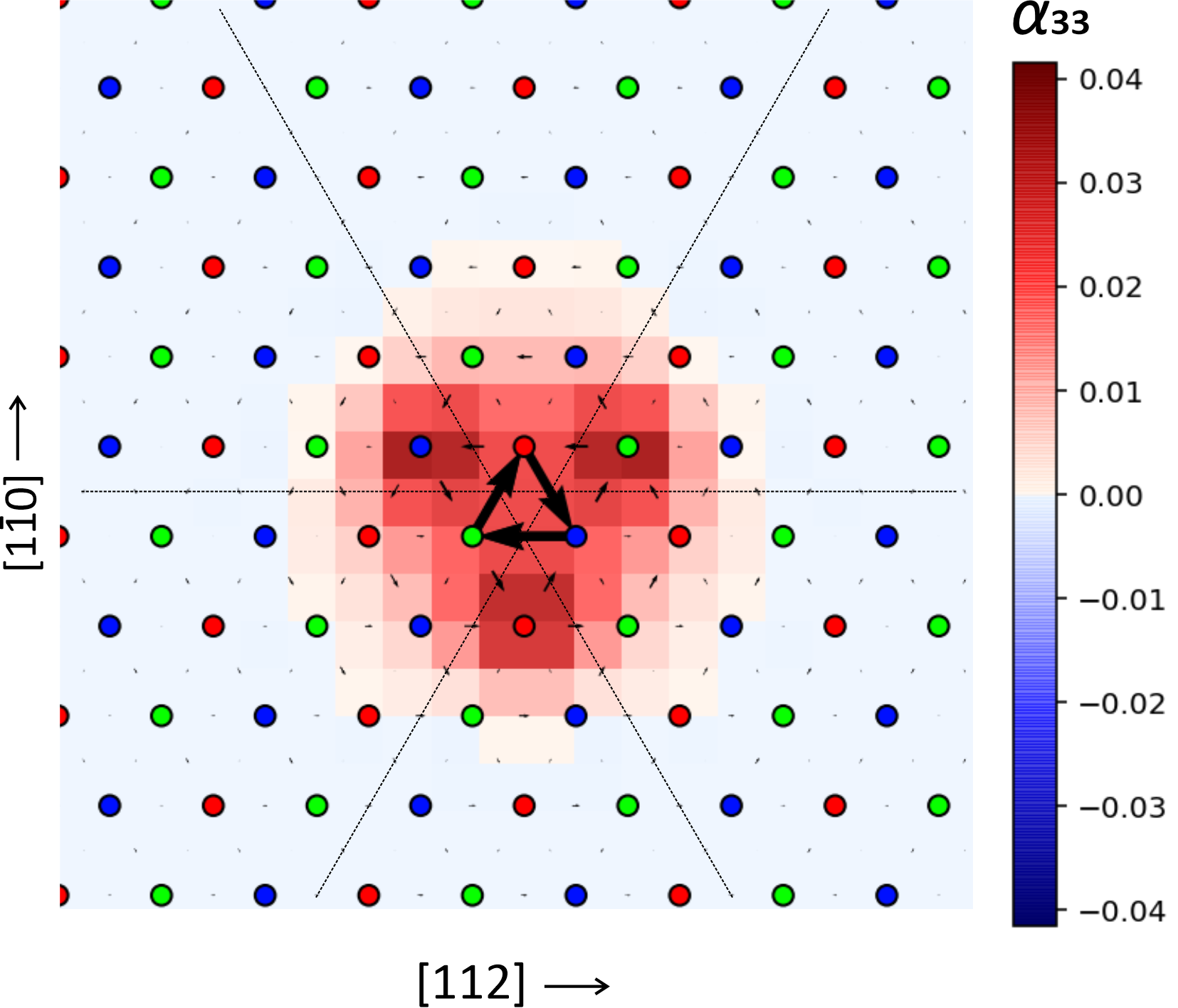}
	\caption{Differential displacement map of the $1/2\langle111\rangle$ screw dislocation in Mo relaxed by our ANN potential. The arrows represent the displacement the atoms and are normalized to $1/3$ of the Burgers vector.
	The blue, red, and green dots represent atoms at the A, B, and C planes along $\langle 111 \rangle$, respectively. The color code represents the $\alpha_{33}$ component of the Nye tensor (in \AA$^{-1}$ units). The Nye tensor is calculated at the atoms and linearly interpolated in-between the atoms. The dashed lines indicate the symmetry equivalent \{110\} planes.}
	\label{fig:screw}
\end{figure}

 We have modeled a screw dislocation using a cylindrical cluster with its axis along $\langle111 \rangle$ and a radius of 220\;\AA. 
The dislocation line coincides with the axis of the cylinder. The dislocation is introduced by mapping the displacement field predicted by anisotropic elasticity theory~\cite{hirth1969dislocation,anderson2017theory} on the positions of the atoms in the cluster. For this, we used the elastic constants calculated by our ANN potential (see Table~\ref{Tabni}). 
Periodic boundary conditions were applied along $\langle111\rangle$. The atoms at an outer shell of 45\;$\mathrm{\AA}$ thickness were kept fixed and all other atoms are relaxed.

Fig.~\ref{fig:screw} shows the differential displacement map of the core structure of the $1/2\langle111\rangle$ screw dislocation. 
As can be seen the displacement field at and around the core spreads symmetrically into the \{110\} planes intersecting the dislocation, in agreement with earlier DFT calculations~\cite{woodward2001ab, shimizu2007first}. Hence, the present potential provides an excellent description of the atomic relaxations in and at the dislocation core. 
The displacement field has the highest values at the three atoms adjacent to the dislocation line, and it decays rapidly to zero within a few lattice constants from the core.
The symmetry of the core is also investigated by evaluating the Nye tensor following the approach proposed by Hartley \textit{et al.}~\cite{hartley2005representation,mendis2006use}. The color map in Fig.~\ref{fig:screw} shows the $\alpha_{33}$ Nye tensor component where the two indices indicate the directions of the dislocation line and of the Burgers vector, respectively. The distribution of the local content of the dislocation shows a similar symmetric pattern to the differential displacement map.
These reveal that the core is non-degenerate. On the other hand, a degenerate screw core is characterized by the asymmetric spreading of the core into three \{110\} planes~\cite{vitek1970core,mendis2006use}.
Nevertheless, the symmetry of the core and the strain field revealed by the present ANN potential is consistent with the picture emerged from DFT based calculation on screw dislocations in bcc materials~\cite{dezerald2014ab} as well as with the row potential model predicted core by Takeguchi \textit{et al.}~\cite{takeuchi1979core}.

%%%%%%%%%%%%%%%%%%%%%%%%
%%%%%%%%%%%%%%%%%%%%%%%%
%%%%%%%%%%%%%%%%%%%%%%%%
\section{Fracture properties}\label{sec:frac}

During fracture, a crack tip is under a multi-axial stress state and several phenomena dominate at different length scales such as bond breaking and lattice trapping, phase transformation, crack tip blunting, and dislocation nucleation and emission~\cite{gumbsch1998controlling,groh2015fracture,wu2015brittle}. Capturing all these phenomena is challenging for almost all classes of interatomic potentials~\cite{groh2015fracture,wu2015brittle,moller2014comparative}. For example, Moller \textit{et al.}~\cite{moller2014comparative} showed that interatomic potentials giving similar surface energies and unstable stacking fault energies predict different crack tip behaviors. As the brittle fracture is correlated to surface energies, and the ductile fracture to unstable stacking fault energies, the different crack tip behaviour may be attributed to the limited transferability of the interatomic potentials.
Therefore, accurate description of the crack tip behaviour under fracture constitutes a challenge for the machine learning based interatomic potentials as well.

\begin{table*} 
	\caption{Critical stress intensity factors K$_\mathrm{Rice}$ and K$_\mathrm{Griffith}$ obtained from LEFM using DFT and ANN potential calculated parameters [see Eqs. (10)-(11)]. K$_\mathrm{1c}$ is obtained from full atomistic simulations of fracture employing the ANN potential. Five different crack configurations in the $\langle111\rangle\{112\}$ and $\langle111\rangle\{110\}$ slip systems are considered. The intensity factors are in  MPa$\sqrt{\mathrm{m}}$. For each crack configuration the angles $\theta$ and $\phi$ between the slip and the crack plane and between the crack front normal and the Burgers vector are listed. The observed behaviour in the atomistic simulations, i.e., brittle vs ductile, is given in  the last column. The bold faced values indicate the lowest of the  K$_\mathrm{Rice}$ and K$_\mathrm{Griffith}$.}\label{tab:kval}
	\begin{ruledtabular}
		\begin{tabular}{cccc|ccc|ccc|cc}
			\multicolumn{4}{c|}{ } & \multicolumn{3}{c|}{DFT} &\multicolumn{3}{c|}{ANN} & \multicolumn{2}{c}{Large scale simulations }\\
			\hline
			Crack direction/ &Slip&  \multirow{2}{*}{$\theta$} & \multirow{2}{*}{$\phi$} &\multirow{2}{*}{K$_\mathrm{Griffith}$} & \multirow{2}{*}{K$_\mathrm{Rice}$} & Expected & \multirow{2}{*}{K$_\mathrm{Griffith}$} & \multirow{2}{*}{K$_\mathrm{Rice}$} & Expected & \multirow{2}{*}{K$_\mathrm{1c}$}  & Observed\\
			plane    &system & & &&  &behaviour & &  &behaviour  &  & behaviour\\
			\hline 
			[010]/(100)&$\langle111\rangle\{110\}$  &45.0 & 35.2 &  \textbf{1.42} &  2.20 &  Brittle & \textbf{1.44} & 2.18 & Brittle&1.16  & Brittle \\
			$[011]$/(100) &$\langle111\rangle\{112\}$  &35.3 & 0 &  \textbf{1.42} &  2.36 &  Brittle & \textbf{1.44} & 2.53 &  Brittle &1.02  & Brittle \\
			$[001]/(110)$ &$\langle111\rangle\{112\}$  &54.7& 0 &\textbf{1.32}& 1.72& Brittle &\textbf{1.34}&1.70&Brittle&1.16 &Brittle\\
			$[\bar{1}10]/(110)$ &$\langle111\rangle\{110\}$ & 90 & 35.4 & \textbf{1.32}& 2.04& Brittle&\textbf{1.34}&2.02&Brittle&1.20&Brittle\\
			$[1\bar{1}0]/(111)$  &$\langle111\rangle\{110\}$& 90 & 0 & \textbf{1.35}& 1.76& Brittle&\textbf{1.40}&1.74&Brittle&1.18&Brittle\\
		\end{tabular}
	\end{ruledtabular}
\end{table*}

In the present study, we focus on the mode I fracture in Mo. We follow two approaches: In the first approach we evaluate the critical stress intensity factor by employing linear elastic fracture mechanics.
Within isotropic elasticity the critical stress intensity factor for dislocation emission in mode I fracture is defined by Rice theory~\cite{rice1992dislocation}:
\begin{equation}
\begin{split}
    %\MoveEqLeft
\mathrm{K}_\mathrm{Rice}= 
(\mathrm{cos}^2(\theta/2)\,\mathrm{sin}(\theta/2))^{-1}\\ \sqrt{\frac{2\mu}{1-\nu}\gamma_{us} \left( 1+(1-\nu)\,\mathrm{tan}^2\phi \right)}\qquad,\label{riceeqn}
  \end{split}
\end{equation}
where $\mu$ and $\nu$ are the shear modulus and the Poisson ratio, respectively. $\theta$ and $\phi$ denote the angle between the slip plane and the crack plane and the angle between the crack front normal and the Burgers vector, respectively. $\gamma_{us}$ is the unstable stacking fault energy which is obtained from the generalized stacking fault energy (GSFE) curve~\cite{vitek1968intrinsic}.

The brittle fracture on the other hand is investigated using Griffith theory which postulates that a pre-existing crack in a material will grow if the elastic energy release during crack growth equals the energy required to cleave surfaces~\cite{griffith1921vi}. Based on this, the critical stress intensity factor, ($\mathrm{K}_\mathrm{Griffith}$), for cleaving two surfaces is written as  
\begin{equation}
\mathrm{K}_\mathrm{Griffith}=\sqrt{\frac{2\,E\,\gamma}{1-\nu^2}}\qquad,
\label{greqn}
\end{equation}
where $\gamma$ is the surface energy, and $E$ is the Young modulus, and $\nu$ is the Poisson ratio.

In order to calculate the critical stress intensity factor, we have considered two slip systems, namely the $\langle111\rangle\{110\}$ and $\langle111\rangle\{112\}$, with five different crack orientations as shown in Table~\ref{tab:kval}. The Rice and Griffith critical intensity factors were calculated using the unstable stacking fault and surface energies of the respective slip plane from DFT and ANN potential calculations. $\theta$ and $\phi$ values are taken from our earlier work such that the combination of these two yields the highest possibility of nucleating dislocations~\cite{groh2015fracture}. The calculated values of the critical stress intensity factors are tabulated in Table~\ref{tab:kval}. 

The DFT calculated critical intensity factors for ductile behaviour, i.e., K$_\mathrm{Rice}$, vary from 1.72\;MPa$\sqrt{\mathrm{m}}$ for the $[\bar{1}10]/(110)$ crack configuration in the $\langle 111\rangle\{110\}$ slip system to 2.36\;MPa$\sqrt{\mathrm{m}}$ for the $[011]/(100)$ crack configuration in the $\langle 111\rangle \{112\}$ slip system. 
The K$_\mathrm{Rice}$ values calculated by the ANN potential are, within less than 0.17\;MPa$\sqrt{\mathrm{m}}$, in excellent agreement with DFT. There is also an excellent agreement between DFT and ANN for the critical stress intensity factors for brittle behaviour (K$_\mathrm{Griffith}$): The ANN values deviate less than 0.05\;MPa$\sqrt{\mathrm{m}}$ from the DFT values. Nevertheless, for all crack configurations considered, K$_\mathrm{Griffith}$ is smaller than K$_\mathrm{Rice}$, i.e., a brittle failure mechanism is predicted by linear elastic fracture mechanics.

\begin{figure}[t]
	\includegraphics[width=\columnwidth]{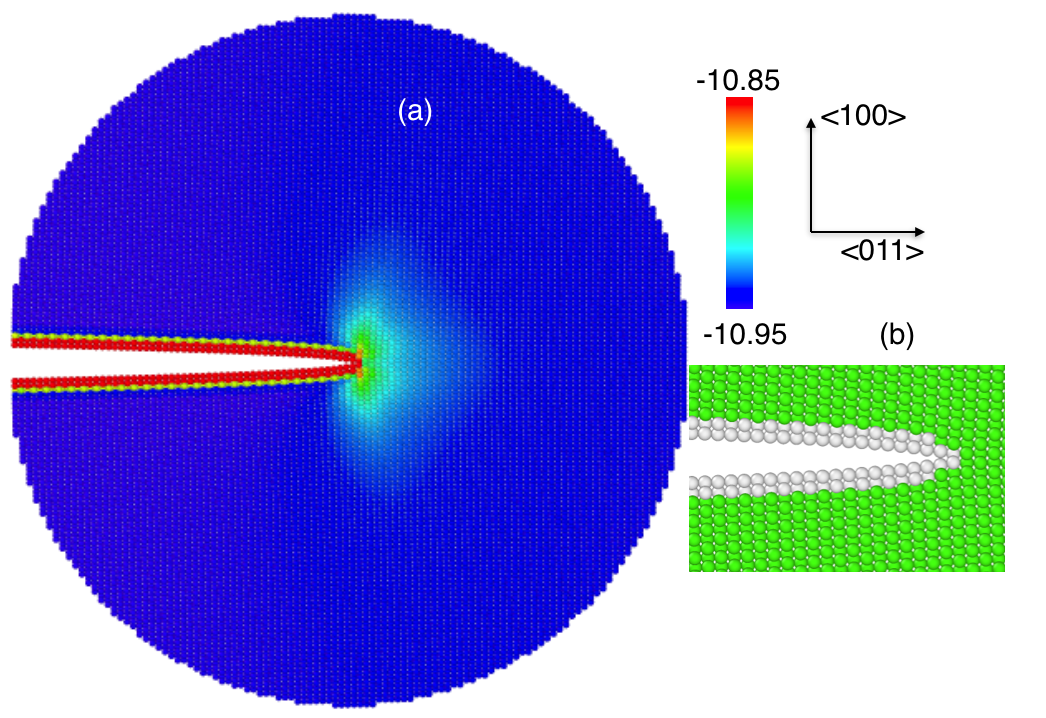}
	\caption{(a) Schematic representation of the crack tip at the
	$\mathrm{K}/\mathrm{K}_\mathrm{1c} $ = 1.0 loading for the  [011]/(100) crack system. The color code represents the energy of the atoms. Cold/hot colors represent lower/higher potential energies, respectively. Red balls are atoms at the bounding surface of the crack. (b) Zoom into the crack tip.   Green balls indicate atoms in the bcc configuration, and white balls atoms in a configuration other than bcc, hcp, fcc, or diamond.
	}
	\label{fig:cractip}
\end{figure}

In the second approach, we explicitly investigate the crack tip opening by employing atomistic calculations. 
We have introduced an atomically sharp crack at the centre of an orthogonal parallelepiped simulation cell by applying the isotropic displacement field $\boldsymbol{u}=(u_1,u_2)$ of mode I fracture under plane strain conditions~\cite{andric2018atomistic}: 
\begin{equation}
\label{dis1}
u_1=\frac{\Delta \mathrm{K}_I}{\mu}\sqrt{\frac{r}{2\pi}}\;cos\frac{\beta}{2}\left(1-2\nu+sin^2\frac{\beta}{2}\right)\qquad,
\end{equation}
\begin{equation}
\label{dis2}
u_2=\frac{\Delta \mathrm{K}_I}{\mu}\sqrt{\frac{r}{2\pi}}\;sin\frac{\beta}{2}\left(2-2\nu+cos^2\frac{\beta}{2}\right)\qquad,
\end{equation}
where $\Delta \mathrm{K}_I$ is the increment in the stress intensity factor and
\begin{equation}
\label{dis3}
r=\sqrt{x_1^2+x_2^2}\qquad,
\end{equation}
and
\begin{equation}
\beta=tan^{-1}\left(\frac{x_2}{x_1}\right)\qquad.
\end{equation}
$x_1$ is along the crack propagation direction and $x_2$ is along the crack plane normal. Periodic boundary conditions are applied along $x_3$.

The crack loading [see Eqs. (\ref{dis1}) and (\ref{dis2})], were applied quasistatically to all atoms in the simulation cell: After atomic relaxation the stress intensity factor was increased by 0.02\;MPa$\sqrt{\mathrm{m}}$.
The positions of atoms within a cylinder of radius 120\;$\mathrm{\AA}$ were relaxed until the change in the total energy was less than $10^{-4}$\;eV or the force on each atom was lower than $10^{-4}$\;eV/$\mathrm{\AA}$. The atoms in an outer shell of thickness 40\;$\mathrm{\AA}$ were kept fixed. This procedure was repeated until the applied stress intensity factor raises above 140\% of the K$_\mathrm{Griffith}$ values derived by linear elastic fracture mechanics [see Eq.~(\ref{greqn}) and Table~\ref{tab:kval}].

Fig.~\ref{fig:cractip}(a) shows a snapshot of the crack tip at the loading stage of $\mathrm{K}/\mathrm{K}_{1c}$=1.0 for the crack system [011]/(100), where K$_{1c}$ is the critical stress intensity factor. 
As can be clearly seen, the crack tip is
sharp and it remained sharp during the complete loading circle. Moreover, the crack advanced by cleavage of the
crack plane, i.e., this crack system is brittle. The same behaviour is observed for all considered crack and slip configurations, i.e., in alignment with the linear elastic fracture mechanics the large scale atomistic calculations reveal brittle fracture.

The critical stress intensity factors calculated by large scale atomistic simulations,  K$_{1c}$, are listed in Table~\ref{tab:kval}. 
K$_{1c}$ varies from 1.02 to 1.2\;MPa$\sqrt{\mathrm{m}}$. The large scale atomistic fracture calculations underestimate the critical stress intensity factor with respect to Griffith's theory for brittle fracture. This discrepancy can be attributed to the isotropic displacement field that was applied to simulate the crack loading.
Nevertheless, the ANN potential provides a consistent description of fracture behavior both in large scale atomistic fracture simulations as well as in terms of Griffith and Rice theories. 
%%%%%%%%%%%%%%%%%%%%%%%%
%%%%%%%%%%%%%%%%%%%%%%%%
%%%%%%%%%%%%%%%%%%%%%%%%
\section{\label{conclu}Conclusions}

In the present study, we have developed a Behler-Parinello–type neural network interatomic potential for Mo. The NN architecture consists of four layers with 33 nodes at the first layer that corresponds to the 15 and 18  radial and angular symmetry functions and 20 nodes at each hidden layer. The NN potential was trained and evaluated on 326 DFT calculated structures that include different bulk crystal phases, vacancies, surfaces and surface decohesion profiles, and GSFEs. The root mean square error, both in the training and validation data sets, was smaller than 5\;meV/atom, with respect to DFT calculations. 

The present potential provides an excellent description of the relative stability of different low energy Mo polymorphs with respect to the ground state bcc phase compared to DFT calculations.
Apart from the high energy phase of diamond, the energies of fcc, hcp, and sc are excellent agreement with DFT (error less than 15\;meV/atom).
The elastic constants of bcc Mo are within 15\% with respect to the DFT calculated values. 
Furthermore, it overcomes limitations of EAM and MEAM potentials in describing surface energies, i.e., it provides and an excellent qualitative and quantitative description of the energies of low index Mo surfaces with respect to DFT calculations. Moreover, the unstable stacking fault energies for both considered slip systems are within 12\% in agreement with DFT calculated values. These properties, i.e., surface and generalized stacking fault energies, play a dominant role in fracture.

We evaluated the transferability of the ANN potential by investigating the properties of the $1/2\langle111\rangle$ screw dislocation core. In agreement with previous DFT studies, our calculations
revealed a non-degenerate, undissociated, and compact core. The strain was found to be spread symmetrically into the \{110\} planes.
We have further employed our ANN potential to study mode I fracture in Mo in the $\langle111\rangle$ $\{111\}$ and $\langle111\rangle$ $\{112\}$ slip systems. In all cases and in agreement with DFT calculations, linear elastic fracture mechanics predict brittle behaviour. 
This has been further confirmed by employing large scale atomistic calculations of fracture
and crack propagation. These calculations revealed that the crack propagates by cleavage of the crack plane and that the crack tip remains sharp throughout the crack propagation during the complete loading cycle. Moreover, no brittle-ductile competition was predicted for all slip systems and crack orientations considered. 

In summary, the present study demonstrates that ANN potentials trained and evaluated using a relatively small but carefully chosen dataset of DFT calculated atomic geometries and total energy differences, outperforms "traditional" EAM or MEAM potentials, provide a near-DFT accuracy of the calculated properties, and are well suited for large scale fracture simulations. 

\appendix
\appendix*
\bibliography{fraref1.bib}
\end{document}